\documentclass[12pt,a4paper,oneside,headings=small]{scrartcl}
\usepackage[utf8]{inputenc}
\usepackage[english]{babel}
\usepackage{graphicx,geometry}
\usepackage{fancyhdr,mathptmx,float}
\usepackage[T1]{fontenc}

\usepackage{helvet,array}
\newcolumntype{L}[1]{>{\raggedright\let\newline\\\arraybackslash\hspace{0pt}}m{#1}}
\usepackage[x11names]{xcolor}
\usepackage{booktabs}
\usepackage{amssymb,amsfonts,amsbsy,amsmath}
\usepackage{bm}
\usepackage{upgreek}
\usepackage{parskip}
\usepackage{subfig}
\usepackage{tikz}
\usepackage{enumitem}
\setlist[itemize]{leftmargin=*} 
\setlist[enumerate]{leftmargin=*} 

\usepackage[comma,authoryear,sort&compress]{natbib}
\fancypagestyle{firstPage}{%
\fancyhead[L]{}
\fancyfoot[L]{{\textit{\fontsize{9}{7}\selectfont 14$^{th}$ International Symposium on Hazards, Prevention and Mitigation of Industrial Explosions\\
Braunschweig, GERMANY - July 11-15, 2022}}}
\fancyfoot[R]{\vspace{-5mm}\includegraphics[trim=2mm 0mm 2mm 0mm,clip=true,width=0.09\textwidth]{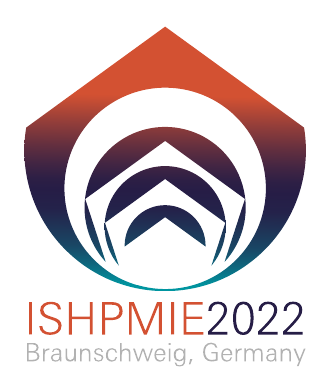}}

}

\setkomafont{section}{\normalfont\bfseries}
\RedeclareSectionCommand[
  beforeskip=-1\baselineskip,
  afterskip=.1\baselineskip]{section}
\setkomafont{subsection}{\normalfont\itshape}
\RedeclareSectionCommand[
  beforeskip=-1\baselineskip,
  afterskip=.1\baselineskip]{section}
\setkomafont{subsubsection}{\normalfont\itshape}
\RedeclareSectionCommand[
  beforeskip=-1\baselineskip,
  afterskip=.1\baselineskip]{section}

\addto\captionsenglish{
        
        }
\usepackage[format=plain,
            labelfont=it,
            textfont=it]{caption}

\begin{document}
\thispagestyle{firstPage}

\begin{center}
\textbf{\LARGE Review on CFD modeling of electrostatic powder charging during pneumatic conveying}\\[15pt]
Holger Grosshans$^{a,b}$, Simon Jantac$^{a}$\\[5pt]
$^a$ Physikalisch-Technische Bundesanstalt (PTB), Braunschweig, Germany\\
$^b$ Otto von Guericke University of Magdeburg, Institute of Apparatus- and Environmental Technology, Magdeburg, Germany\\[5pt]
E-mail: {\color{blue} \textit{holger.grosshans@ptb.de}}\\[5pt]
\end{center}

\section*{Abstract}
Thus far, Computational Fluid Dynamics (CFD) simulations fail to reliably predict the electrostatic charging of powder during pneumatic conveying.
The lack of a predictive tool is one reason for unwanted discharges and growing deposits that make a plant a prime candidate for an explosion.
This paper reviews the numerical models' state-of-the-art, limitations, and progress in recent years.
In particular, the discussion includes the condenser model, which is up to today most popular in CFD simulations of powder flow electrification but fails to predict most of its features.
New experiments led to advanced models, such as the non-uniform charge model, which resolves the local distribution of charge on non-conductive particle surfaces.
Further, models relying on the surface state theory predicted bipolar charging of polydisperse particles made of the same material.
Whereas these models were usually implemented in CFD tools using an Eulerian-Lagrangian strategy, powder charging was recently successfully described in an Eulerian framework.
The Eulerian framework is computationally efficient when handling complete powders;
thus, this research can pave the way from academic studies to simulating powder processing units.
Overall, even though CFD models for powder flow charging improved, major hurdles toward a predictive tool remain.

\medskip
\noindent
\textbf{Keywords:} \textit{Simulation, electrostatics, pneumatic conveying, industrial explosions}

\section{Introduction}


One way to control powder charging would be to analyze an industrial process by simulations.
Then, based on the results, one could adapt the facility's design or choose its operating parameters to limit the generating charge.
However, the simulation of the charging of flowing powder is extremely challenging.
It requires coupling the equations of fluid mechanics (turbulent conveying airflow), surface science (triboelectric charge exchange, adhesion), and electromagnetism (electrostatic attraction of charged particles).
Each of these scientific sub-fields being complex by itself, their numerical coupling of these equations is yet more difficult.
For some of the mentioned physical processes, the mathematical equations are not even clear to date.

In particular, particles change their charge through various physical mechanisms: 
through ionized gas or dissipation, but most often through contact with other surfaces. 
The lacking understanding of the physics and chemistry of particle charging explains the limited success of related numerical model formulation.
These models usually require heavy tuning of parameters, or the predicted charge differs from experimental measurements by several orders of magnitude.
For these reasons, CFD simulations are not mature enough to reliably evaluate the charging of particulates during processing.

\begin{figure}[tb]
\centering
\includegraphics[trim=0cm 0cm 0cm 0cm,clip=true,width=0.55\textwidth]{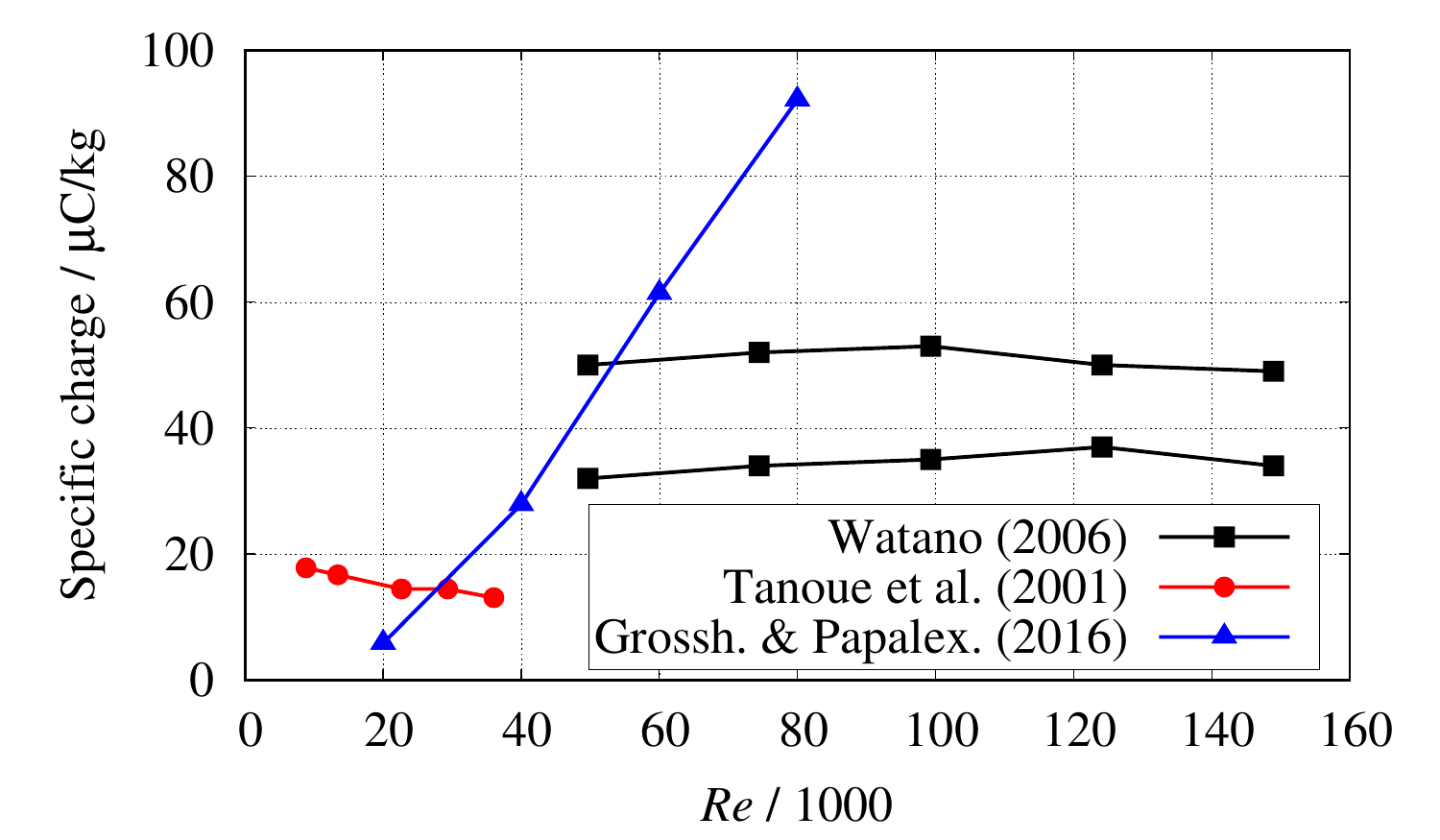}
\caption{CFD simulations of the powder charge after pneumatic conveying depending on the flow Reynolds number.}
\label{fig:qre}
\end{figure}

Figure~\ref{fig:qre} compiles CFD predictions of powder charging by three groups using different codes.
Whereas \citet{Tan01} predicts the powder charge to decrease with increasing Reynolds number, the data of \cite{Gro16a} suggests the opposite.
According to \cite{Wat06}, the Reynolds number has nearly no influence.
Even though each group simulated different particle material and sizes, the contradictory trend of the results is surprising since the flow Reynolds number is the dominating operation condition of pneumatic conveying.

It is emphasized that powder flow electrification is not simply the sum of the charging of the individual particles.
Instead, fluid dynamics, electrostatics, and triboelectricity give rise to complex intertwined interactions, e.g.:
\begin{itemize}
\item The dynamics of a particle-laden flow determines the frequency and severeness of particle/surface and particle/particle contacts and, thus, the charge accumulation of powder~\citep{Gro17a,Jin17}.
\item The charge exchange during one contact does not only depend on the charge carried by the particle itself, but also on the electrostatic field generated by the charges of all other present particles and induced charges on surfaces~\citep{Mat95c,Mat97}.
\item The electrostatic field significantly changes the powder flow pattern through electric forces and, thus, alters the dynamics of subsequent contacts \citep{Dho91}.
\end{itemize}
These interactions cause perplexing phenomena, such as particles moving counter to the main gas flow due to the emerging electrostatic field~\citep{Myl87}.
In other words, only having a correct particle charging model is not enough for a correct prediction of powder charging.
In essence, the hazard of electrostatic charge accumulation to the operational safety of an industrial facility must be evaluated at a powder flow level.

This paper reviews the state-of-the-art, limitations, and progress in recent years of the numerical modeling of electrostatic charging of powder flows.
Out of all industrial powder operations, pneumatic conveying, due to the high flow velocities, leads by far to the highest charge levels~\citep{Kli18}.
Therefore, this review focuses on simulations of pneumatic conveying.
Nevertheless, the research questions in pneumatic powder conveying are often similar to those of closely related fields, and their model development stimulates each other.
In particular, this review summarizes advances in simulations, purely experimental studies are only included if they directly led to a model.
Otherwise, the reader is referred to the reviews of \citet{Lacks19} on general triboelectricity, of \citet{Chow21} on single particle charging models, of \citet{Mat10} on experimental electrostatics, of \citet{Mehr17} on charging in fluidized beds, and of \citet{Wong15} on charging in pharmaceutics.

This paper is organized as follows:
Sections~2 to~4 present the available numerical concepts to model the flow of charged powder in pneumatic conveying.
More specifically, Section~2 gives an overview of the methods to simulate the carrier gas phase.
Section~3 provides an outline of the different methods to simulate the dynamics of powder, including approaches to compute the electric field and the electrostatic forces on the particles.
Section~4 summarizes the models of triboelectric charging on a single particle level.
The final section gives the author's opinion on the future perspectives of the field.

\section{Modeling the turbulent carrier gas flow}
\label{sec:gas}

Given that particles collect most of their charge during contacts, and contacts are driven by aerodynamic forces, the simulation of the carrier gas flow plays a paramount role in powder charging.
The gas flow in pneumatic conveyors is described by the Navier-Stokes equations.
That means by the mass and momentum balance of incompressible Newtonian fluids in Eulerian framework,
\begin{subequations}
\begin{equation}
\label{eq:mass}
\nabla \cdot {\bm u} = 0
\end{equation}
\begin{equation}
\label{eq:mom}
\frac{\partial {\bm u}}{\partial t} + ({\bm u} \cdot \nabla) {\bm u}
= - \frac{1}{\rho} \nabla p  + \nu \nabla^2 {\bm u} + {\bm F}_\mathrm{s}\, ,
\end{equation}
\end{subequations}
where ${\bm u}$ denotes the fluid's velocity, $p$ its pressure, $\rho$ its density, $\nu$ its kinematic viscosity, and $t$ the temporal coordinate.
The source term ${\bm F}_\mathrm{s}$ accounts for the momentum transfer from the particles to the carrier fluid.
Both equations rely on a fundamental physical principle, Eq.~(\ref{eq:mass}) on the assumption that mass can neither be created nor destroyed and Eq.~(\ref{eq:mom}) on Newton's second law of motion extended to fluids.
Since analytical solutions were found only for a few simple flow cases, solving the above equations requires numerical simulations.

Most of the time, pneumatic conveyors operate at high Reynolds numbers, which means in fully turbulent mode.
The most exact method to simulate turbulence, termed \textit{direct numerical simulation}~(DNS), resolves all length- and time-scales of fluid motion on the numerical grid.
However, turbulent flows of a high Reynolds number exhibit a wide range of scales.
Resolving all spatial and temporal scales requires a fine grid and a small time-step, resulting in a high computational effort.
Therefore, when simulating pneumatic powder conveying, turbulence is usually modeled instead of resolved.

In early computations of powder charging, not even the mean flow was solved but approximated by an analytical velocity profile.
Afterward, the first simulations appeared using the \textit{Reynolds-averaged Navier-Stokes}~(RANS) approach~\citep{Kol89,Tan99,Tan01}.
In RANS, equations~(\ref{eq:mass}) and~(\ref{eq:mom}) are temporally or ensemble-averaged.
Due to the averaging new unclosed terms arise, the so-called Reynolds stresses.
Widespread closures include the mixing-length model~\citep{Bal78} and the standard $k-\epsilon$~\citep{Jon72} and $k-\omega$~\citep{Wil98} models.

In other words, the RANS approach solves the mean flow but models all turbulence scales.
This is reasonable when only time-averaged quantities are of interest rather than turbulent fluctuations.
However, powder receives most of its charge when the particles reflect on the conveying duct's walls.
Especially near-wall turbulence drives these impacts' frequency, velocity, and angle.
Thus, the turbulence model’s deficiency directly impairs the prediction of powder charging by RANS simulations.

For several years, \textit{Large Eddy Simulation}~(LES) of powder flow charging has been feasible~\citep{Kor14,Gro16a}.
LES computes the filtered governing equations~(\ref{eq:mass}) and~(\ref{eq:mom});
only the turbulent motions larger than the filter size are resolved on the grid.
Similar to RANS, new unclosed terms corresponding to the small (subfilter) scales appear through the filter operation.
The rationale of LES stems from Kolmogorov's hypothesis that the small-scale structures are universal and can, thus, be modeled.
Some of the most popular closures include the \citet{Sma63} model, the dynamic approach by \citet{Ger91}, the scale similarity model by \citet{Bar80}, and the implicit approach by \citet{Bor92}.
They all approximate the sub-filter terms from the resolved flow field, even though experiments showed the correlation is weak~\citep{Liu94}.

The computational effort of LES is much higher compared to RANS.
But if the grid resolution is fine enough, a considerable part of the turbulence energy spectrum is resolved.
Then, the influence of the turbulence model diminishes, and LES becomes exact.
LES is especially reliable if the ratio of the characteristic particle to flow time is high.
The ratio of the characteristic particle to flow time, which is the particle's Stokes number,
\begin{equation}
St=\dfrac{\tau_\mathrm{p}}{\tau_\mathrm{f}},
\end{equation}
determines the dynamics of the air-particle interaction.
For those particles of a high Stokes number, inertial forces act as a high-pass filter.
Their trajectories are influenced by large-scale but not by small-scale turbulence.
Thus, the requirement to the grid resolution relaxes when simulating the charging of high Stokes number particles.

\begin{figure}[tb]
\centering
\subfloat[]{\includegraphics[trim=0mm -15mm 120mm 0mm,clip=true,width=.48\textwidth]{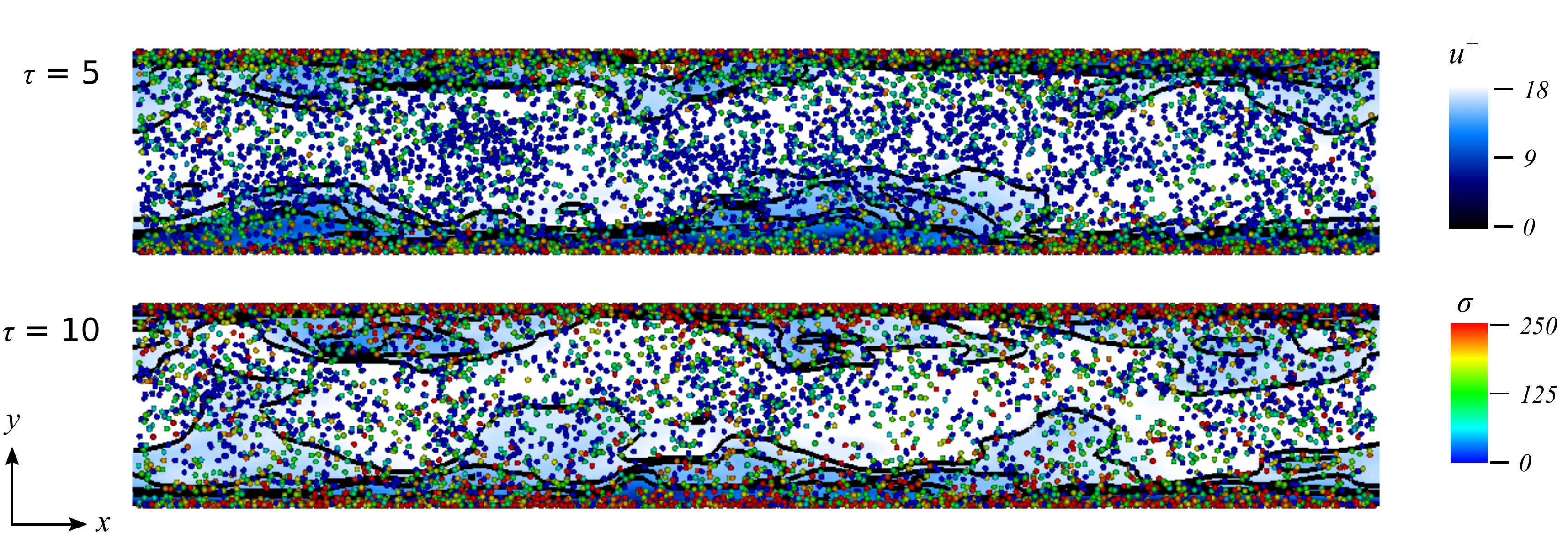}%
\label{fig:plates}%
}
\quad
\subfloat[]{
\begin{tikzpicture}[scale=1]
\node at (0,0) {\includegraphics[trim=0cm 0cm 0cm 0cm,clip=true,width=.45\textwidth]{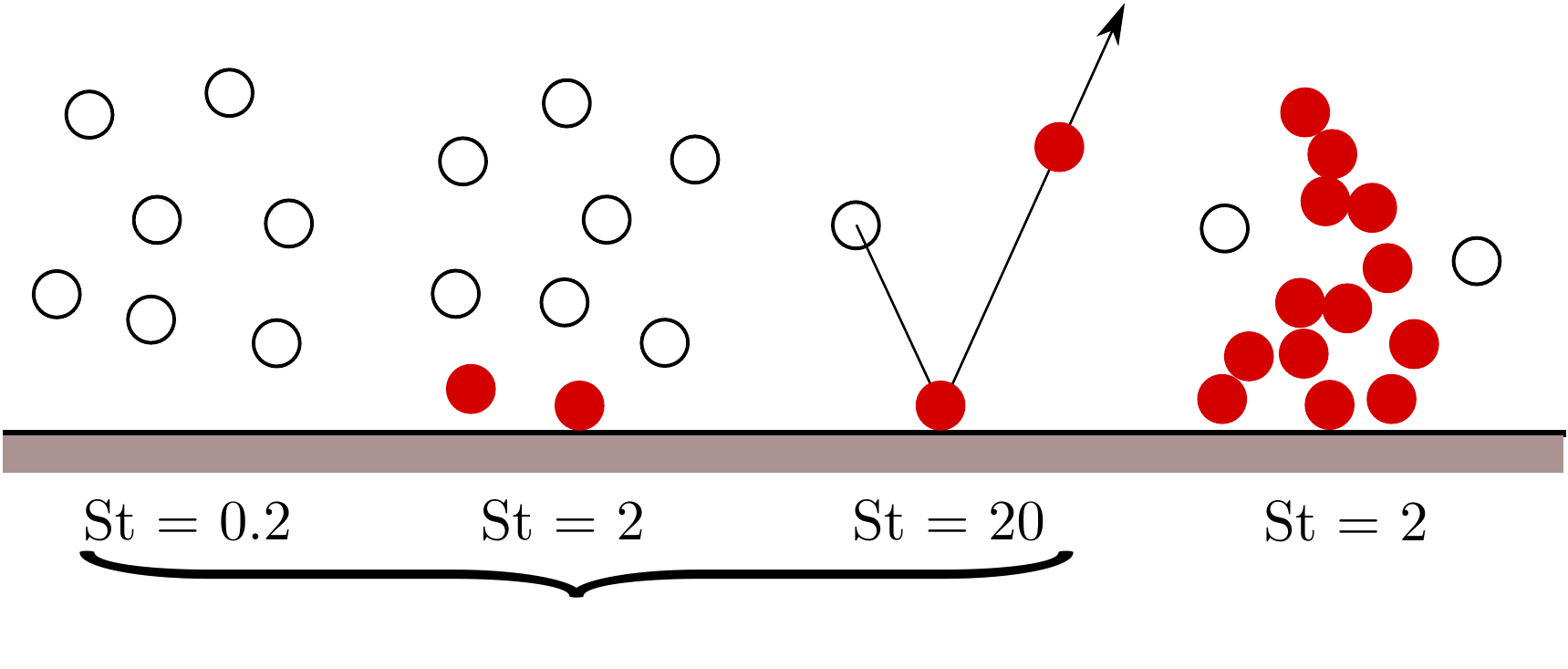}};
\draw [->,>=latex,ultra thick] (-3.0,2.0) node [above right,black,font=\footnotesize,xshift=0] {$u_\mathrm{gas}$} -- (-2.0,2.0);
\draw [->,>=latex,ultra thick,blue] (.8,2.0) node [above,align=center,black,font=\footnotesize,xshift=-15] {particle-bound\\charge transport} -- (.8,1.3);
\draw [->,>=latex,ultra thick,blue] (2.9,2.0) node [above,align=center,black,font=\footnotesize] {inter-particle\\charge diffusion} -- (2.9,1.3);
\node at (-.9,-1.8) [above,font=\footnotesize] {low $\phi$};
\node at (3.0,-1.8) [above,font=\footnotesize] {high $\phi$};
\end{tikzpicture}%
\label{fig:mech}}
\caption{DNS of powder electrification in a channel flow depending on the Stokes number ($St$) and particle volume fraction ($\phi$).
(a) is for $St=$~20, the colors indicate the particles' charge.
(adapted with permission from \citet{Gro17a,Gro18c})}
\end{figure}
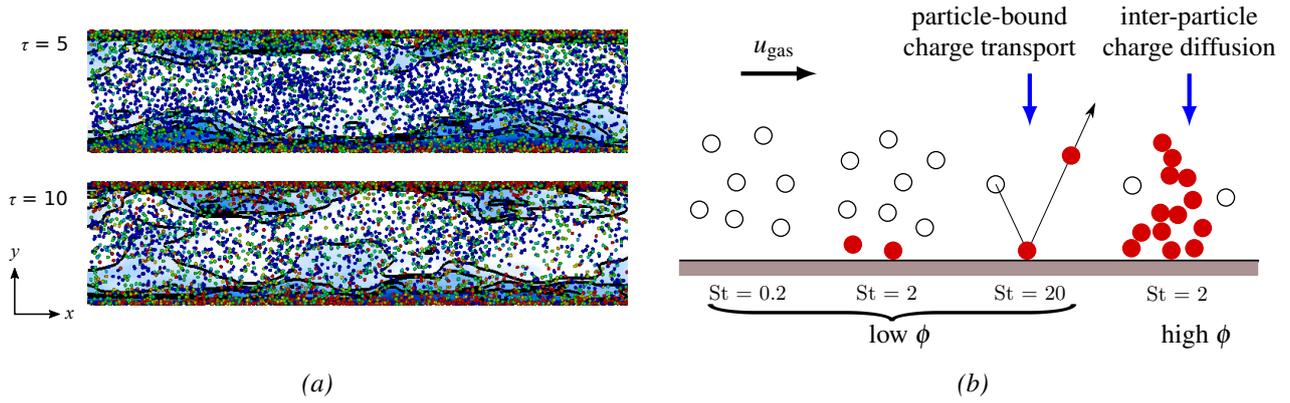

Only recently, the first DNS of electrifying powder flow was achieved~\citep{Gro17a}.
However, DNS can not simulate complete industrial unit operations.
Instead, it is limited to generic domains and low Reynolds numbers, such as the channel flow of a friction Reynolds number of~360 in Fig.~\ref{fig:mech}.
These DNS revealed, at a previously unknown level of detail, the small-scale mechanisms that determine the powder charging rate.
More precisely, the mechanisms sketched in Fig.~\ref{fig:mech} dominate the charge transfer from the channel walls to and within the powder flow:
\textit{particle-bound charge transport} for highly inertial particles and \textit{inter-particle charge diffusion} for low inertial particles in case of high particle volume fractions ($\phi$).
Identifying these mechanisms implies the possibility to control the electrification of powder flows by imposing flow conditions that purposely trigger these mechanisms.


\section{Modeling electrostatically charged powder flow}
\label{sec:powder}

Contrary to the carrier gas, which is continuous, the powder forms a dispersed phase.
Powder consists of abundant particles.
The amount of particles an their related solid/gas interface area restrict the choice of the numerical method.
Those numerical methods for multiphase flows that resolve the phase interface on the grid, such as volumes-of-fluids, level-set, or marker-and-cell, are computationally too expensive.
Instead, pneumatic conveying is usually modelled by the Eulerian-Lagrangian or the Eulerian-Eulerian approach.
In both appoaches, the carrier gas is described in the Eulerian framework, as discussed in Sec.~\ref{sec:gas}.
The particulate phase is either described in Lagrangian framework, that means each particle is tracked individually, or in the Eulerian framwork, where the powder is modelled as a continuum.

\subsection{Lagrangian}


Most simulations of powder charging during pneumatic conveying use the Lagrangian framework to describe the particle flow.
In the Lagrangian framework, each particle is treated individually as a point-mass whose motion is computed as 
\begin{equation}
\label{eq:dropacc}
m_\mathrm{p}\frac{{\mathrm{d}} {\bm u}_{\mathrm{p}}}{{\mathrm{d}} t} = \sum {\bm F} \, ,
\end{equation}
where ${\bm u}_{\mathrm{p}}$ is the velocity and $m_\mathrm{p}$ the mass of the given particle.
The term on the right-hand side represents the sum of all specific external forces acting on the particle which are elaborated in the following sub-section.

The advantage of the Lagrangian approach is that there is no limitation on $St$ and polydispersity can be handled more easily compared to the Eulerian approach.
However, the ratio of the average particle diameter to the characteristic flow scale is assumed to be low.
Further, the numerical coupling of Lagrangian particles to the carrier phase poses a challenge.

The computational effort of the Lagrangian approach scales with the number of particles, $N$.
Some sub-models scale linearly with $N$ whereas others, such as collisions between particles, require the comparison of particle pairs.
The computational effort of comparing particle pairs scales by $O(N^2)$.
Advanced algorithm reduce the cost, for example, Fast Multipole Methods (FMM)~\citep{Rok90} to $O(N\log N)$.
Nevertheless, operations that require evaluating particle pairs remain elaborative.
Especially for pneumatic conveying systems, which consist of missions of particles, these operations can easily inflate the overall computational time.
Therefore, the models describing Lagrangian particles have to be carefully chosen to optimize the equation system's accuracy and efficiency.

Further, the Lagrangian framework is limited to study the transport through one pipe instead of a complete pneumatic system, and for dilute or pulsed conveying where the particle number is low.
Or for academical research, looking at fundamental charging methods in only a section of the complete pipe.
For fundamental research, the Lagrangian approach plays out its strength, namely the resolution of individual particle trajectories.

\subsection{Forces on a particle}

The specific external forces acting on a particle are given by
\begin{equation}
\label{eq:dropacc2}
\sum {\bm F} =  {\bm F}_{\mathrm{g}} + {\bm F}_{\mathrm{coll}} + {\bm F}_{\mathrm{ad}} + {\bm F}_{\mathrm{vdW}} + {\bm F}_{\mathrm{el}} \, ,
\end{equation}
where ${\bm F}_{\mathrm{g}}$ denotes the gravitational, ${\bm F}_{\mathrm{coll}}$ the collisional, ${\bm F}_{\mathrm{ad}}$ the drag,  ${\bm F}_{\mathrm{vdW}}$ the van der Waals, and ${\bm F}_{\mathrm{el}}$ the electric field forces acting on the particle.

The selection of forces included in the simulation model depends on the specific conveying system under consideration:
for vertical conveying of high Stokes number particles, the particle dynamics with and without gravity is nearly identical~\citep{Marc07};
thus, the gravitation can be neglected.
For horizontal conveying of low Stokes number particles, gravity determines the particles' trajectories and, thus, their charging.
Therefore, gravitation is considered in all simulations of horizontal conveying.

The specific collisional force term ${\bm F}_{\mathrm{coll}}$ accounts for both inter-particle and particle-wall collisions.
Collisions between particles requires the comparison of particle pairs, which is, as discussed above, computationally expensive.
Therefore, inter-particle collisions are neglected whenever possible.
During dilute conveying, particles collide seldom with each other~\citep{Elg94}.
Therefore, inter-particle collisions are usually only modelled when simulating dense conveying.

Due to the high flow velocities, the aerodynamic drag acting on a particle~\citep{Som12},
\begin{equation}
\label{eq:ffl}
{\bm F}_\mathrm{ad} = - \frac{\pi}{2} C_\mathrm{d} \, \rho \, r^2_\mathrm{p} \, \vert {\bm u}_\mathrm{rel} \vert \,{\bm u}_\mathrm{rel} \, ,
\end{equation}
is part of all pneumatic conveying simulations.
In this equation, ${\bm u}_{\mathrm{rel}}$ the particle velocity relative to the gas, and $C_{\mathrm{d}}$ is the particle drag coefficient.
The drag coefficient is computed according to the relation provided by \citet{Sch33} as a function of the particle Reynolds number,
\begin{align}
\label{eq:drag}
C_\mathrm{d} = \dfrac{4}{Re_\mathrm{p}} \left(6 + Re_\mathrm{p}^{2/3} \right)
\quad \text{with} \quad
Re_\mathrm{p} = 2 \vert {\bm u}_\mathrm{rel} \vert \, r_\mathrm{p} / \nu \, .
\end{align}
Originally, this expression was derived experimentally for idealized conditions, namely for isolated, spherical particles exposed to an undisturbed airflow.
These idealizations generally do not hold for pneumatic conveying.
Many new drag correlations were proposed in the recent years, reflecting non-spherical particles~\citep{Wac12}, shear flow due to the pipe's walls~\citep{Zeng09}, or the disturbance of the flow by nearby particles~\citep{Krav19,Tang14}.
.
However, the dynamics of charged particles is different from uncharged ones and so is their drag.
The drag correlation for charged particles have, with the exception of the thesis of \citet{Ozl22}, not been researched yet.
Given that the near-wall dynamics of particles determines their charging during pneumatic conveying, choosing a suitable drag correlation is decisive for predicting powder flow charging.

There are other aerodynamic forces (besides drag) acting on a particle, summed up by the Basset-Boussinesq-Oseen~(BBO) equation~\citep{Max83}.
These include the virtual mass force that is required to drag along the surrounding fluid when the particle is accelerated.
The virtual mass force is important for the case of a low solid-fluid density ratio which is not typical for pneumatic transport.
The effect of a non-uniform flow around a particle is accounted for by the Faxen force.
Further, the Saffman force is caused by the rotation of a particle due to large velocity gradients in shear flows.
Both Faxen and Saffman forces, vanish if the particle size is small compared to the scale of the local flow gradients.
The assumption of non-rotating particles also allows to neglect the Magnus force.
The time delay in building up a boundary layer in the vicinity of the particles' surface is described by the Basset history term. 

Also, the aerodynamic drag imposes a force on the fluid phase which is given by ${\bm F}_{\mathrm{s}}$ in equation~(\ref{eq:mom}).
Once again, for dilute conveying, where the number of particles is low, ${\bm F}_{\mathrm{s}}$ can be neglected.

Van der Waals forces can be stronger than gravitational forces if the particles are small \citep{Tom09}.
For airborne particles during pneumatic conveying, van der Waals forces play no role.
They act only during a minuscule duration when the distance in-between particles or a particle and a wall is of the nanometer order, therefore, the particle's momentum change is negligible.
Nevertheless, van der Waals forces can form dust deposits on the surfaces of pipes or other components.
Thus, for the prediction of deposits, van der Waals forces need to be considered \citet{Gro19e}.


Finally, the last term in Eq.~(\ref{eq:dropacc2}) describes the electrostatic force acting on a particle that carries the charge $Q$,
\begin{equation}
\label{eq:fel}
{\bm F}_{\mathrm{el}} = Q \, {\bm E} \, ,
\end{equation}
which can dominate the dynamics of particles in pneumatic conveyors.
The electric field strength, ${\bm E}$, is given by Gauss' law,
\begin{equation}
\label{eq:gauss}
\nabla \cdot {\bm E} = \dfrac{\rho_{\mathrm{el}}}{\varepsilon_0} \, ,
\end{equation}
where $\varepsilon_0$ is the electrical permittivity and the electric charge density, $\rho_{\mathrm{el}}$, reflects the charge carried by all particles in the system.
Gauss' law involves only $O(N)$ operations and is, therefore, fast to solve.
However, an extremely fine grid is required to resolve the gradient of the electric field caused by charged particles in close proximity.

Assuming the charge of each particle is located at its centre point, a mathematical equivalent formulation to Eq.~(\ref{eq:gauss}) is Coulomb's law,
\begin{equation}
\label{eq:coulomb}
{\bm E}_m = \sum\limits_{n=1,n\neq m}^N \dfrac{Q_n \, {\bm z}_{n,m}}{4 \, \pi \, \varepsilon_0 \, |{\bm z}_{n,m}|^{3}} \, .
\end{equation}
Herein, ${\bm E}_m$ is the electric field at the position of particle $m$, $N$ the number of all particles in the system, and ${\bm z}_{n,m}$ a vector pointing from the centre of particle $n$ to the centre of particle $m$.

Equation~(\ref{eq:coulomb}) contains only Lagrangian variables and, therefore, requires no grid to solve.
Drawback compared to Eq.~(\ref{eq:gauss}) is that it involves comparisons of particle pairs, thus, $O(N^2)$ operations.

Similar solutions to this problem were independently proposed by \citet{Kol16} and \citet{Gro17e}, combining the numerical advantages of Gauss' and Coulomb's law.
More specifically, their hybrid approaches superimpose the far-field interactions computed with Eq.~({eq:gauss}) and the Coulombic interactions between the particle and its neighboring particles.
This approach is both fast and accurate and generally recommended for future simulations.
In particular, it is more suitable for wall-bounded flows than the Ewald summation or the P$^3$M method \citep{Yao18}.

Nevertheless, the point charge assumption impedes the prediction of particle dynamics resulting from inhomogeneous charge distribution on the particles' surface.
For example, the attraction of particles of the same polarity due to induced charges~\citep{Qin16} cannot be captured.
For fluidized beds, \citet{Kol18b} recently included particle polarization due to surrounding charges.
The development of advanced numerical models reflecting the surface charge distribution is expected to boost the accuracy of future pneumatic conveying simulations.

\subsection{Eulerian}

The Eulerian-Lagrangian approach suits especially numerical studies of laboratory-scaled systems.
But even the expense of $O(N)$ operations limits the number of particles that can be computed simultaneously.
Contrary, the description of powder in Eulerian framework opens the possibility to handle complete technical flows consisting of a vast amount of particles.
In Eulerian description, the powder is treated as a continuum whose properties are averaged in each computational cell.

While the Eulerian-Eulerian approach is popular for general powder flow simulations, only recently a few studies appeared where it was employed to the charge generation of particle-laden flows.
\citet{Kol18} developed a two-fluid model including the effect of electrostatic forces on the particles and charge diffusion through the random motion of particles.
\citet{Ray18} and \citet{mont20} developed new formulations to compute electrostatic charging of particles in Eulerian framework.
Whereas the mentioned works are limited to mono-disperse particle size distributions, \citet{Ray20} expanded their earlier model to bi-disperse granular flows.
Finally, \citet{Gro20h} presented a description for the transport of charged poly-disperse powder in Eulerian framework using the direct quadrature method of moments~(DQMOM)~\citep{Mar05}.

All these Eulerian formulations are steps toward the simulation of the charge build-up in technical flow facilities.
Nevertheless, the accuracy of these models lacks way behind Lagrangian formulations.

\section{Particle charging models}

All methods discussed in the previous section to simulate pneumatic powder conveying assume the particles to be smaller than the cells of the computational mesh.
In other words, the numerical grid does not resolve the gas-solid interfaces.
Thus, all physical processes taking place on the particles' surface need to be modeled explicitly.
These processes include, for example, aerodynamic drag, heat and mass transfer, collisions, phase change, adhesion, and chemical reactions.
Often the underlying physics of these processes is complex and sometimes not even understood yet.
Complex physical mechanisms needs to be simplified to obtain a computational efficient model suitable for CFD simulations.
Usually, the uncertainty of the particle models defines the leading error to the overall simulation model.

This section reviews CFD models for the electrostatic charge transfer between a particle and an object.
The implementation in a CFD approach requires the model to be accurate, computationally efficient to handle a vast amount of particles, able to predict charge transfer based on the data available in a CFD framework, and valid for conditions relevant to technical flows.
These requirements impede the usage of detailed theoretical approaches, such as quantum mechanical or atomistic calculations~\citep{Fu17}.

For more than five decades, the most spread CFD model to predict particle contact charging is the so-called \textit{condenser model}~\citep{Soo71,Mas76,John80}.
Its name refers to the analogy of particle charging to the temporal response of a capacitor (also known as a condenser) in a resistor-capacitor (R-C) circuit.
Even though the condenser model appeared over the years in different variants, all formulations base on the same assumptions:
\begin{enumerate}
\item A particle charges upon contact with another surface.
\item The driving force for the charge transfer is the contact potential difference of the material pair, $V$, and the charge held by the particle before contact.
\item The polarity of the transferred charge is always the same.
\item The amount of transferred charge depends on the electrical properties and the contact kinematics.
\item The particle charge saturates asymptotically.
\end{enumerate}

Thus, during collisions of two particles of the same material, which is the typical situation for particles being part of the same powder batch, no charge transfers because their contact potential is the same.
Nevertheless, charge may exchange if at least one of the two particles carries a charge prior to the contact.
In the original formulation by \citet{Soo71}, the charge transfer between two particles, $\Delta Q_n = -\Delta Q_m$, during the collision contact time, $\Delta t_{\mathrm{p}}$, reads
\begin{equation}
\label{eq:chargeexpp}
\Delta Q_n = \dfrac{C_n C_m}{C_n+C_m} \left( \dfrac{Q_m}{C_m} - \dfrac{Q_n}{C_n} \right) \left( 1- \mathrm{e}^{-\Delta t_{\mathrm{p}} / \tau_{\mathrm{p}}} \right) = -\Delta Q_m \, .
\end{equation}
In the above equation, $C_n$ and $C_m$ denote the capacity of both particles and $\tau_{\mathrm{p}}$ their charge relaxation time.

Afterward, \citet{John80} expanded the model to the impact of a spherical particle with a plane surface such as a wall or a plate.
In opposite to particle-particle collisions, in this situation, the two objects in contact are usually of dissimilar material.
Thus, the total impact charge from the target to the particle, $\Delta Q$, is given by the sum of the dynamic charge transfer to the particle caused by the contact potential, $\Delta Q_{\mathrm{c}}$, and the transferred pre-charge, $\Delta Q_{\mathrm{t}}$, i.e.,
\begin{equation}
\label{eq:chargeextot}
\Delta Q = \Delta Q_{\mathrm{c}} + \Delta Q_{\mathrm{t}} \, .
\end{equation}
The dynamic charge transfer during the wall-particle contact time $\Delta t_{\mathrm{pw}}$ is, as for a parallel plate condenser, given by
\begin{equation}
\label{eq:chargeex}
\Delta Q_{\mathrm{c}} = - C V \left( 1- \mathrm{e}^{-\Delta t_{\mathrm{pw}} / \tau_{\mathrm{pw}}} \right)
\end{equation}
where $C$ is the electrical capacity and $\tau_{\mathrm{pw}}$ the charge relaxation time.

It is commonly assumed \citep{John80,Kol89} that the pre-charge is distributed uniformly on the particles' surface, $A_\mathrm{p}$.
Further, if the charge within the particle-target contact area, $A_{\mathrm{pw}}$, is completely transferred, $\Delta Q_{\mathrm{t}}$ equals
\begin{equation}
\label{eq:preuni}
\Delta Q_{\mathrm{t}} =  - \frac{A_{\mathrm{pw}}}{A_{\mathrm{p}}}  Q_n \, .
\end{equation}

Even though this concept holds only for the transfer of electrons during the contact of conductors, is was often successfully applied to the charging of insulators by assigning an effective work function~\citep{Chow18}.

As mentioned above, the condenser model went through some evolutionary steps, one being the refinement of the contact potential difference to~\citep{Mat00} 
\begin{equation}
V = V_\mathrm{c} - V_\mathrm{e} -  V_\mathrm{b} + V_\mathrm{ex} \, .
\end{equation}
Therein, the total contact potential difference is separated into contributions by the surface work functions~($V_\mathrm{c}$), the image charge~($V_\mathrm{e}$), the space charge by surrounding charged particles~($V_\mathrm{b}$), and other external electric fields~($V_\mathrm{ex}$).

\begin{figure}[tb]
\centering
\subfloat[]{\includegraphics[trim=0mm -15mm 0mm 0mm,clip=true,width=.22\textwidth]{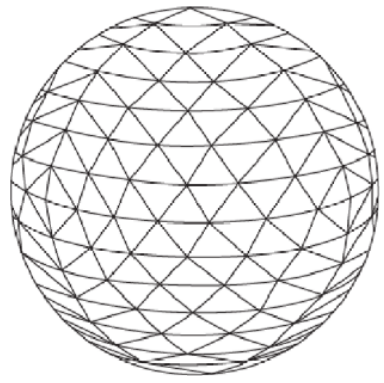}%
\label{fig:Yos03}%
}
\qquad
\subfloat[]{
\includegraphics[trim=0cm 0cm 0cm 0cm,clip=true,width=.62\textwidth]{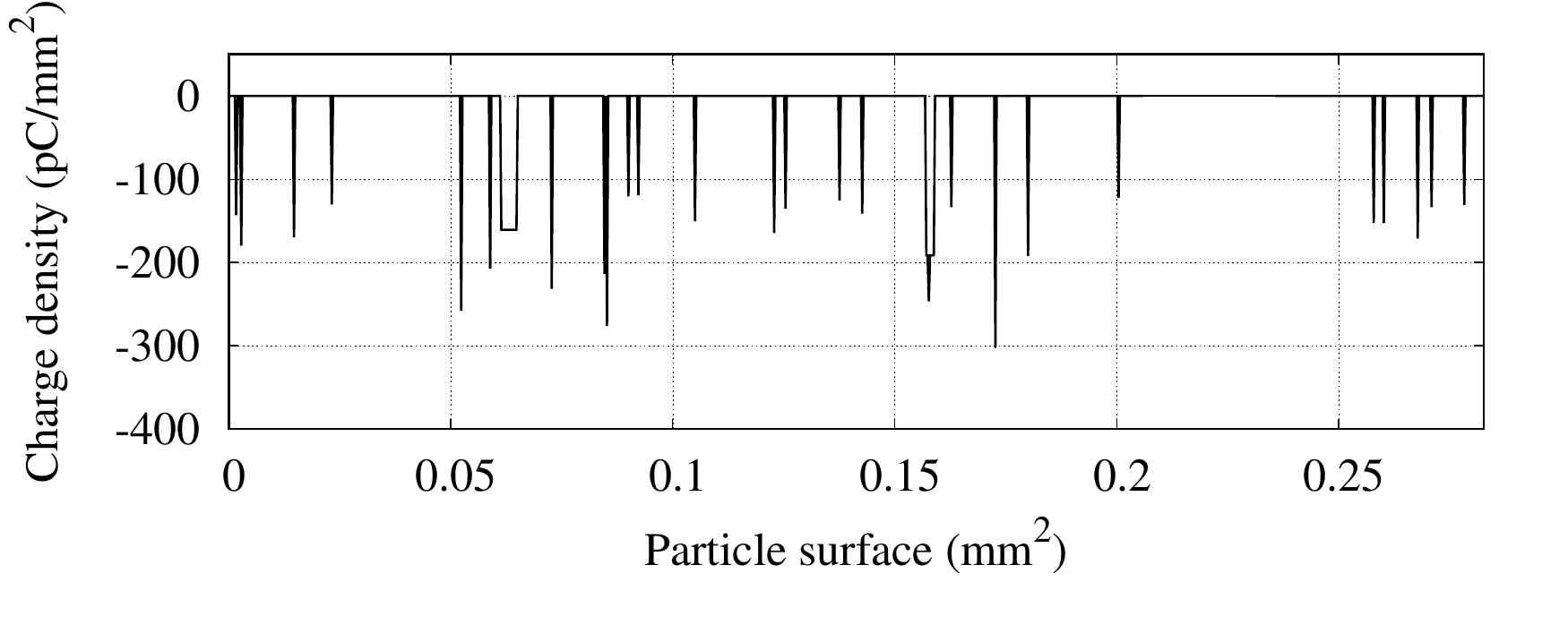}
\label{fig:Gro16f}}
\caption{(a) Charging site concept of \citet{Yos03}. (b) Resolved charge density on a particle's surface after pneumatic conveying \citep{Gro16f}.}
\end{figure}

\begin{figure}[tb]
\begin{minipage}[c]{0.47\textwidth}
\vspace{-\ht\strutbox}
\centering
\includegraphics[trim=0mm 0mm 0mm 0mm,clip=true,width=0.75\textwidth]{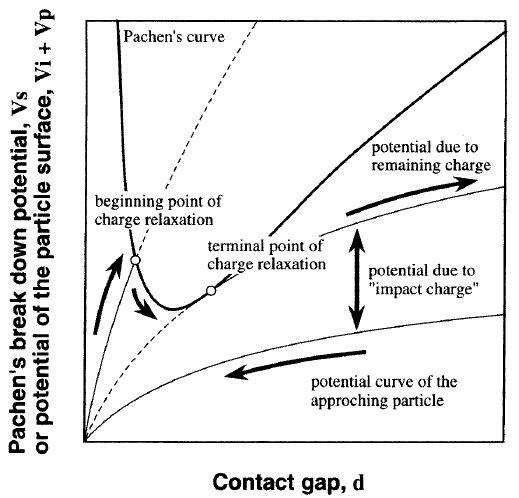}
\end{minipage}
\qquad
\begin{minipage}[c]{0.47\textwidth}
\vspace{-\ht\strutbox}
\vspace{4mm}
\centering
\includegraphics[trim=0cm 0cm 0cm 0cm,clip=true,width=1.0\textwidth]{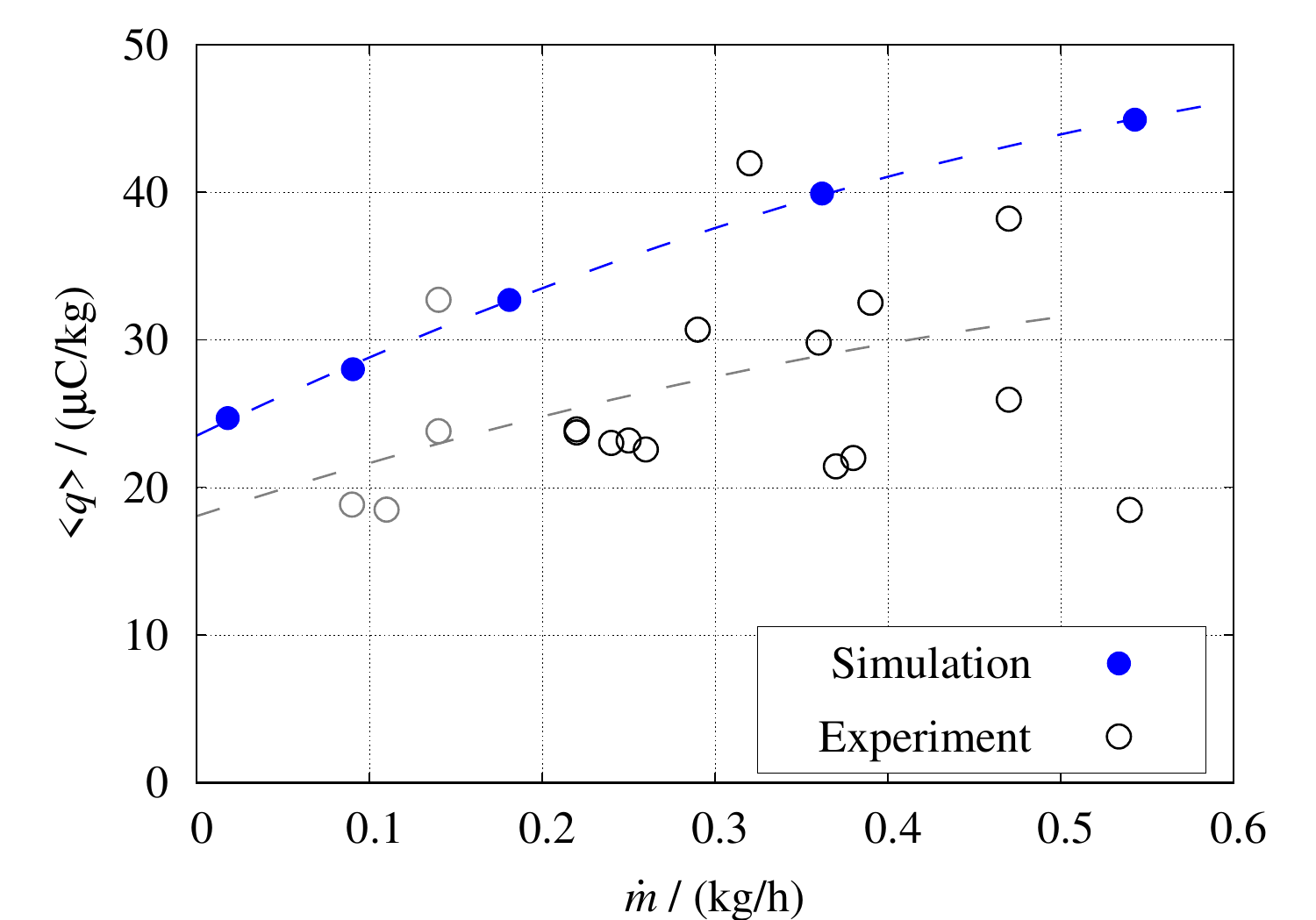}
\end{minipage}

\bigskip
\begin{minipage}[c]{0.47\textwidth}
\vspace{-\ht\strutbox}
\caption[]{Charge relaxation model \citep{Mat95c}.}
\label{fig:Mat97}
\end{minipage}
\qquad
\begin{minipage}[c]{0.47\textwidth}
\vspace{-\ht\strutbox}
\centering
\caption{CFD simulations of the electrification of PMMA particles during conveying using an empirical charging model.}
\label{fig:Gro21c}
\end{minipage}
\end{figure}

The above formulations of the condenser model assume a uniform charge distribution on the particles surface.
However, charge does not distribute uniformly on insulative surfaces, such as polymers.
Therefore, the strong scatter of the impact charge in the single-particle experiments of \citet{Mat03} was attributed to a non-uniform charge distribution on the particle's surface.

As response to the observed scatter, several models resolve the charge location on particle surfaces.
\citet{Yos03} introduced the concept of charging sites which take up charge individually, see Fig.~\ref{fig:Yos03}.
Using the concept of charging sites, \citep{Gro16f} extended the condenser model to the \textit{non-uniform charge model} for particle/surface and inter-particle collisions.
This formulation leads to a wide range of possible outcomes of a contact event, which partially explains the scatter of the experimentally measured charging behavior of a single PTFE particle.
The non-uniform charge model was used to simulate pneumatic powder transport.
Figure~\ref{fig:Gro16f} shows the resolved charge on the particle surface after leaving the duct.
Each peak is caused by an impact.
Some peaks even overlap each other, which means the particle impacted at a location of a charge spot left by a previous impact.

Another group of charging models relies on the surface state theory~\citep{Low86a,Low86b}.
According to it, electrons with high energy levels exist only at the surface of insulators and can transfer to empty surface states of another insulator upon contact driven by their different effective work functions.
These models aim to explain the charging of particles made of the same material.
The low-density limit was recently utilized in models~\citep{Duff08}, in a probabilistic version~\citep{Lacks07}, and in a more general formulation considering the transfer of any charged species~\citep{Kon17}.
By assuming the transfer of charge carriers from one particle to another until they are depleted, the results of this model agreed with two trends in observed in powder flows:
particles charge stronger in highly poly-disperse systems, and big particles are usually positively and small particles negatively charged.

More a charge limitation than a generation model is the \textit{charge relaxation model} \citep{Mat95c} whose principle is visualized in Fig.~\ref{fig:Mat97}.
Therein, the arrows present the evolution of the potential difference between the particle and the wall, which increases after contact.
Discharge takes place at the contact gap where the potential difference equals the gaseous breakdown limit potential, which is given by Paschen's law.
Thus, this model limits the predicted charge exchange. 

Finally, a purely empirical charging model was recently proposed by \citet{Gro21c} for spherical PMMA particles.
The model bases on data from single-particle experiments using the precise same particles as in the simulations.
The CFD simulations agree well with experiments, see Fig.~\ref{fig:Gro21c}, for 200~$\upmu$m particles, but fail for 100~$\upmu$m particles.

However, this model, just as all above-discussed charging models, handles only very specific situations.
A generally predictive charging model that satisfies the requirements of a CFD tool is not in reach yet.
Thus, in the foreseeable future, the particle charging model will remain the largest contributor to the overall error of CFD simulations of powder flow electrification.








\section{Perspectives for future research}

Due to its outstanding complexity, the CFD simulation of powder electrification fails so far.
It requires the solution of an interdisciplinary mathematical model describing turbulence, electrostatics, and triboelectric charging.
This paper reviewed the state-of-the-art and pinpointed toward the future research necessary to improve the numerical predictions.
Highly-resolved direct numerical simulations of the carrier gas flow combined with Lagrangrian simulations of the particle dynamics offer insight in the detailed mechanics of powder charging.
Understanding the dependence of powder charging rate on the conveyors operating parameters, such as velocity or powder mass flow rate, can guide the design of future, safe conveying systems.
The largest contributor to the error of current simulations is the particle charging model.
A generally valid, predictive model seems currently out of reach.
But new single-particle experiments that deliver impact data tailored to pneumatic conveying can improve the accuracy of models for specific particles.
Finally, recent Eulerian-Eulerian formulations open a way to the simulation of powder charging in complete flow facilities.
The next step is to improve the handling of Euler-Euler models of polydisperse particle size distributions.


\section*{Acknowledgements}
This project has received funding from the European Research Council~(ERC) under the European Union’s Horizon 2020 research and innovation programme~(grant agreement No.~947606 PowFEct).


{\setlength{\bibsep}{1pt}
\bibliography{\string~/essentials/publications}}

\end{document}